\documentclass[aps,prl,twocolumn,nofootinbib,
superscriptaddress,
]{revtex4-1}

\usepackage{hyperref}
\usepackage{url}
\usepackage{setspace}
\usepackage{amsmath,amssymb,amscd,braket}
\usepackage{amsfonts}
\usepackage{color}
\usepackage{graphicx}
\usepackage{xcolor}

\newcommand{\im}{\mathfrak{Im}}

\newtheorem{theorem}{Theorem}
\newtheorem{corollary}{Corollary}[theorem]

\newcommand{\bea}{\begin{eqnarray}}
\newcommand{\eea}{\end{eqnarray}}
\newcommand{\be}{\begin{equation}}
\newcommand{\ee}{\end{equation}}
\newcommand{\ba}{\begin{align}}
\newcommand{\ea}{\end{align}}

\begin{document}
\title{
Rigorous bounds on transport from causality
}

\author{Michal P. Heller} \email{michal.p.heller@ugent.be}
\affiliation{Department of Physics and Astronomy, Ghent University, 9000 Ghent, Belgium}

\author{Alexandre Serantes}
\email{alexandre.serantes@ub.edu}
\affiliation{Departament de Física Quàntica i Astrofísica, Institut de Ciències del Cosmos (ICCUB), Facultat de Física, Universitat de Barcelona, Martí i Franquès 1, E-08028 Barcelona, Spain}

\author{Micha\l\ Spali\'nski}
\email{michal.spalinski@ncbj.gov.pl}
\affiliation{National Centre for Nuclear Research, 02-093 Warsaw, Poland}
\affiliation{Physics Department, University of Bia{\l}ystok, 15-245 Bia\l ystok, Poland}

\author{Benjamin Withers}
\email{b.s.withers@soton.ac.uk}
\affiliation{Mathematical Sciences and STAG Research Centre, University of Southampton, Highfield, Southampton SO17 1BJ, UK}

\begin{abstract}
We use causality to derive a number of simple and universal constraints on dispersion relations, which describe the location of singularities of retarded two-point functions in relativistic quantum field theories. We prove that all causal dissipative dispersion relations have a finite radius of convergence. We then give two-sided bounds on all transport coefficients in units of this radius, including an upper bound on diffusivity. 
\end{abstract}

\maketitle

\section{Introduction and results} 
In real-time linear response of relativistic quantum field theory, the singularities of retarded two-point functions are of considerable interest. At finite temperature, they describe collective excitations which include slowly varying hydrodynamic modes as well as transient nonhydrodynamic ones. In cases where holography provides a means of calculating the retarded thermal two-point functions, these singularities are simple poles which encode the spectrum of black brane quasinormal modes (QNMs)
\cite{Birmingham:2001pj,Son:2002sd}. 
The location of the singularities is described by dispersion relations, $\omega(k)$. A study of their analytic structure was initiated in \cite{Withers:2018srf}, using holography. The goal of this work is to explore the constraints on the analytic structure of $\omega(k)$ which follow from causality.

Consider the retarded two-point functions of local operators in Minkowski space,
\be
G^R(x,y) = -i \theta(x^0 - y^0)\left<\left[{\cal O}(x),{\cal O}(y)\right]\right>.
\ee
$G^R(0,x)$ is only non-zero for $x$ in the past closed light cone, $x\in \overline{V}_-$; this follows from the $\theta$-function and the commutation of ${\cal O}$ at spacelike-separated points.
In addition we take $G^R(x,y)$ to be a tempered distribution.\footnote{Note that an instability would violate this condition and so we restrict our analysis to only (linearly) stable phases.} Both of these conditions work together to dictate analyticity properties of its Fourier transform,
\be
\tilde{G}^{R}(p) = \int d^{d}x G^R(0,x) e^{i p\cdot x}.\label{Gtilde}
\ee
Specifically, provided $\im(p)$ is inside the open forward light cone, $\im(p) \in V_+$, then $e^{i p\cdot x}$ acts to exponentially suppress the integrand \eqref{Gtilde}. This, combined with $G^R(0,x)$ being a tempered distribution, means the integral and its derivatives converge there and thus $\tilde{G}^{R}(p)$ is analytic \cite{StreaterAndWightman, Haag:1992hx}. Thus $\tilde{G}^{R}(p)$ is analytic for $\im\, \omega > |\im\, k|$ where $p^\mu = (\omega, k,0,0,\ldots)^\mu$ for a rotationally invariant state. 
Hence, any singularity of $\tilde{G}^{R}(p)$, including poles and branch points, whose location is described by the complex function $\omega(k)$, necessarily obeys
\be
\boxed{\im\,\omega(k) \leq |\im\, k|.}\label{cond}
\ee
Here and throughout we have set the light cone speed to $c=1$, but it can be reinstated at any point on dimensional grounds. 
Using only \eqref{cond}, which is a necessary condition for causality, we prove a number properties of $\omega(k)$ of physical significance for both hydrodynamic and nonhydrodynamic modes of quantum field theories.
Note that \eqref{cond} also applies to any model, not necessarily a microscopic quantum field theory, provided its retarded Green's functions are causal and tempered.

Our main result is as follows:

\begin{theorem}[Bounds on transport]\label{thm:bestbound}
Let $\omega(k)$ be analytic in a disk centred at $k=0$ with radius $R$, with Taylor series $\omega(k) = \sum_{n=0}^\infty a_n k^n$. The causality constraint \eqref{cond} implies for $n>0$,
\be
|a_n| \leq \frac{2(n^2-1) - (1+(-1)^n)\sin\theta_n}{n^2-1}\frac{2}{\pi R^{n-1}} - \frac{2 \im(a_0)}{R^n},\label{bestbound}
\ee
and also $\im(a_0) \leq 0$, where $\theta_n = \arg(a_n)$.
\end{theorem}

The analyticity assumption around $k=0$ holds when fluctuations are negligible such as for large-$N$ QFTs.
As in \cite{Withers:2018srf, Heller:2020uuy}, we regard the radius of convergence of the series appearing in this bound, $R$, as an intrinsic computable microscopic scale of the system under consideration. Then, \eqref{bestbound} provides nontrivial two-sided bounds on all transport coefficients in units of this intrinsic microscopic scale. We have not been able to show that the right hand side of \eqref{bestbound} is sharp; there may be a smaller coefficient we can write down on the right hand side following only from \eqref{cond}.

By taking $R \to \infty$ it follows from Theorem \ref{thm:bestbound} that
 
\begin{corollary}\label{cor:entire}
If $\omega(k)$ is an entire function obeying \eqref{cond} then $\omega(k)$ is a polynomial of at most degree one. 
\end{corollary}
For instance, the well-known result that the heat equation is not causal follows from Corollary \ref{cor:entire} since there, $\omega(k) = -i D k^2$, which is entire. Also, the QNMs of BTZ black branes are entire functions and linear in $k$ \cite{Birmingham:2001pj, Son:2002sd}, consistent with this result. 
Another example are the logarithmic branch points of scalar glueball operators in free $SU(N)$ Yang-Mills theory which have $\omega(k)$ linear in $k$ \cite{Hartnoll:2005ju}, respecting Corollary \ref{cor:entire}.
Of course, one may find higher-order polynomial $\omega(k)$ for modes of \emph{nonrelativistic} theories like Navier-Stokes or large-$D$ gravity \cite{Emparan:2016sjk, Andrade:2018zeb} which do not respect a finite $c$ lightcone.

In the hydrodynamic context, Corollary \ref{cor:entire} demonstrates that the only causal and entire dispersion relation is that which follows from the perfect fluid equations of motion.  It follows that any attempt to modify the dispersion relation by adding, for example, dissipative transport coefficients such as viscosity necessarily results in a complex singularity of $\omega(k)$ and a finite radius of convergence. It has been shown on a case-by-case basis that the radius of convergence of hydrodynamic modes (and QNMs generally) is finite in relativistic theories \cite{Withers:2018srf, Grozdanov:2019kge, Grozdanov:2019uhi, Abbasi:2020ykq, Jansen:2020hfd, Baggioli:2020loj, Arean:2020eus, Heller:2020jif, Heller:2020hnq, Abbasi:2020xli, Asadi:2021hds, Wu:2021mkk, Baggioli:2021ujk, Grozdanov:2021gzh, Jeong:2021zsv, Donos:2021pkk, Huh:2021ppg, Liu:2021qmt, Cartwright:2021qpp}. Our main result, Theorem \ref{thm:bestbound}, proves that it always is, in complete generality.

Having established that complex singularities are an unavoidable feature for all but the simplest of cases, it is natural to ask whether \eqref{cond} constrains which singularities are allowed. Indeed, we show: 

\begin{theorem}[No poles or essential singularities]
If $\omega(k)$ obeys \eqref{cond} then it does not contain any poles or essential singularities.
\end{theorem}

The absence of poles in hydrodynamic dispersion relations was argued in \cite{Heller:2020uuy}. 
We have not considered any nonisolated singularities in this work.
However, we note that branch points are not ruled out by \eqref{cond} and are allowed to appear in a causal $\omega(k)$. Indeed, as pointed out in \cite{Withers:2018srf}, there are causal theories for which the radius of convergence of $\omega(k)$ is set by a square-root branch point.

The bound \eqref{bestbound} is our most general, applying to all coefficients. However, the more information available about structure of a given mode, the more the bound following from \eqref{cond} can be refined. We consider such refined bounds two physical cases of interest, sound and diffusive modes in classical hydrodynamics:

\begin{theorem}[Sound modes]
Let $\omega(k)$ be a sound mode, i.e. in a Taylor series around $k=0$ we have $\omega(k) = v k -i \frac{\Gamma_s}{2} k^2 +\ldots$ with a radius of convergence $R>0$ and where $v, \Gamma_s\in \mathbb{R}$. Then \eqref{cond} constrains
\be
|v| \leq 1,\qquad 0 \leq  \frac{\Gamma_s}{2} \leq \frac{16}{3\pi}\frac{1}{R},\label{sound}
\ee
as well as a bound relating the two coefficients,
\be
\frac{\Gamma_s}{2} \leq 4\sqrt{\frac{2}{3}}\left(\frac{8}{\sqrt{6}\pi} - |v|\right)\frac{1}{R}.\label{soundrel}
\ee
\end{theorem}
Note that for CFT$_d$ one has $|v| = \frac{1}{\sqrt{d-1}}$ and hence \eqref{soundrel} gives upper bounds on $\Gamma_s R$ which are stronger than \eqref{sound} in dimensions $d =2,3,4$. An approximate upper bound on $\Gamma_s$ in terms of $1-v$ was given in \cite{Delacretaz:2021ufg}.

\begin{theorem}[Diffusive modes]\label{thm:diffusion}
Let $\omega(k)$ be a diffusive mode, i.e. in a Taylor series around $k=0$ we have $\omega(k) = -i D k^2 +\ldots$ with a radius of convergence $R>0$ and where $D\in \mathbb{R}$. Then \eqref{cond} constrains
\be
0 \leq  D \leq \frac{16}{3\pi}\frac{1}{R}.\label{diffusion}
\ee
\end{theorem}
For example, in the case of the telegrapher's equation, 
which describes shear channel perturbations in MIS theory ~\cite{Israel:1976tn,Israel:1979wp}, the maximum value of the diffusion constant is set by the relation $DR = 1/2$. Of the examples we have checked in the literature, the largest value of $DR$ we have found is $DR \simeq 0.78$ which occurs for the shear mode the dual of Reissner-Nordstr\"om-AdS$_5$ \cite{Jansen:2020hfd}, at the first `cusp' shown in figure 1 there.\footnote{Using analytically known branch points for RN-AdS$_{d+1}$ \cite{Withers:2018srf, Jansen:2020hfd} which set $R$ at intermediate $T$, one can prove that an additional branch point must appear at high $T$ so that \eqref{diffusion} is not violated. This indeed happens \cite{Jansen:2020hfd} resulting in this cusp.} We also provide a mathematical test function which has $DR\simeq 1.22$ \eqref{testfunction}.
Note that, in theories where $\varepsilon + p = Ts$, \eqref{diffusion} gives an upper bound on $\frac{\eta}{s} \leq \frac{16}{3\pi}\frac{T}{R}$.

In \cite{Hartman:2017hhp} arguments were made for a parametric upper bound on $D$, given approximately by a local equilibration timescale, $\tau_\text{eq}$. A precise version of this bound could be derived from Theorem \ref{thm:diffusion} should one identify $R^{-1}$ with $\tau_\text{eq}$ (as a reminder, we have set $c=1$ throughout).

In the remainder of this Letter we give proofs of each of the theorems above and give some concluding remarks.

\section{Proofs}

\paragraph{Bounds on transport.}
Our proof of Theorem \ref{thm:bestbound} starts by following the derivation of a real-part theorem in \cite{Holland} (the improvement of theorem 3.1 presented there), with adjustments to accommodate the more specific bound at hand, \eqref{cond}.\footnote{Alternative bounds can be constructed using the Borel-Carath\'eodory theorem and associated real-part theorems \cite{Kresin2007SharpRT}, but typically these bounds end up relying only on $\im(\omega)\leq |k|$ and are weaker than the one we derive here, since $|\im(k)|\leq |k|$.} Let $k=r e^{i \theta}$ and for $r < R$ write,
\be
\omega(k) = \sum_{n=0}^\infty a_n k^n = U(r,\theta)+i V(r,\theta).
\ee
Let $a_n = \alpha_n + i \beta_n$. Then
\be
V(r,\theta) =\sum_{n=0}^\infty r^n \left(\alpha_n \sin(n\theta) + \beta_n \cos(n\theta)\right).\label{angular}
\ee
Hence for $n>0$, let $\theta_n = \arg(a_n)$
\bea
|a_n|r^n 
&=& (\alpha_n + i\beta_n)e^{-i \theta_n}r^n\\
&=& \frac{1}{\pi}\int_0^{2\pi} V(r,\theta)\sin(n\theta+\theta_n) d\theta, \label{norm_coeff}
\eea
where we used sine/cosine orthogonality integrals to extract $\alpha_n, \beta_n$ from \eqref{angular}, and we also used that the left hand side is real. The bound \eqref{cond} is $V \leq r |\sin\theta|$, but to bound an integral we need to make sure that what multiplies it is non-negative. Hence instead consider
\bea
|a_n|r^n+2 \beta_0 &=& \frac{1}{\pi}\int_0^{2\pi} V(r,\theta)\left(1+\sin(n\theta+
\theta_n)\right) d\theta,\hspace{1.9em}\\
&\leq & \frac{r}{\pi}\int_0^{2\pi} |\sin\theta|\left(1+\sin(n\theta+\theta_n)\right) d\theta,\\
&=& \frac{2r}{\pi} \frac{2(n^2-1) - (1+(-1)^n)\sin\theta_n}{n^2-1}.
\eea
Sending $r\to R$ we obtain \eqref{bestbound}. Note if $n$ is odd, then the dependence on $\theta_n$ drops out. If $n$ is even and we don't know $\theta_n$ we can obtain an agnostic bound by maximising the right hand side over $\theta_n$, i.e. by picking $\theta_n = -\pi/2$. Finally we note that \eqref{cond} at $k=0$ implies $\beta_0 \leq 0$. 

\paragraph{No poles or essential singularities.}
Given an isolated singularity at $a\in \mathbb{C}$ we can form the Laurent series valid in the punctured disk, $0 < |k-a| < R$
\be
\omega = \sum_{n=n_{\text{min}}}^\infty c_n (k-a)^n,
\ee
for some order of the pole, $n_{\text{min}}<0$ with $c_{n_\text{min}} \neq 0$.
Take $k = a + r e^{i \theta_k}$ then 
\be
\im\, \omega = \sum_{n=n_{\text{min}}}^\infty |c_n| r^n \sin(\arg(c_n)+n \theta_k),
\ee
and in the limit $k\to a$ we have, restricting $\arg(a)$ to its principal value,
\be
|\im\, k| = |a| \sin(|\arg(a)|).
\ee
Thus if we approach $k=a$ at an angle $\theta_k = -(\arg(c_n)-|\arg(a)|)/n_\text{min}$ then we violate \eqref{cond}. Hence \eqref{cond} rules out poles. If we take $n_{\text{min}}\to -\infty$ this rules out essential singularities too, with the limit taken along $\theta_k = 0$.

\paragraph{Sound modes.}
Consider a sound mode $\omega(k) = v k - i \frac{\Gamma_s}{2}k^2 +\ldots $ with $v, \Gamma_s \in \mathbb{R}$. 
Take \eqref{cond} with $k = r e^{i\theta}$ then 
\be
v r \sin(\theta) +  \ldots \leq r
\ee
hence by dividing by $r$ and taking a limit $r\to 0$ we find $|v|\leq 1$, i.e. no superluminal sound modes. Note that using \eqref{bestbound} at $n=1$ gives $|v|\leq 4/\pi$ which does not improve this bound. Next consider $\im\, k = 0$ so that 
\be
-\frac{\Gamma_s}{2} k^2 + \ldots \leq 0.
\ee
By dividing by $k^2$ and taking a limit $k\to 0$ we find $\Gamma_s \geq 0$. Note that using \eqref{bestbound} at $n=2, \theta_2 = \pi/2$ gives $\frac{\Gamma_s}{2}\geq -\frac{8}{3\pi}\frac{1}{R}$ which does not improve this lower bound. To obtain an upper bound, we use \eqref{bestbound} at $n=2, \theta_2 = -\pi/2$, so that $\frac{\Gamma_s}{2} \leq \frac{16}{3\pi}\frac{1}{R}$. Altogether we have the two-sided bound \eqref{sound}. 

To prove the relational bound \eqref{soundrel}, we take the hydrodynamic sound mode with $v \geq 0$, and consider the identity
\begin{equation}
4\delta v r + \frac{\Gamma_s}{2} r^2 = \frac{1}{\pi}\int_0^{2\pi} V(r,\theta)\left(\gamma {+} 4\delta \sin(\theta) {-} \cos(2\theta)\right)d\theta,    
\end{equation}
which follows from \eqref{norm_coeff} upon taking into account that $V(0,\theta)=0$ and $v,~\Gamma_s \geq 0$. We assume that $\delta \in [0,1)$. Under this condition, the function 
\begin{equation}
f(\theta)\equiv\gamma {+} 4\delta \sin(\theta) {-} \cos(2\theta)    
\end{equation}
has absolute minima at $\sin(\theta) = - \delta$, with value 
$\gamma-1-2\delta^2$. Choosing $\gamma = 1 + 2 \delta^2$ ensures that $f(\theta)$ is a non-negative function, thus allowing to employ the bound \eqref{cond}, $V \leq r |\sin(\theta)|$, to obtain the following inequality 
\begin{equation}
\frac{\Gamma_s}{2} \leq 4\delta\left(\frac{4+6\delta^2}{3\pi\delta} - v\right)\frac{1}{R}. 
\end{equation}
The value $\delta = \sqrt{\frac{2}{3}} < 1$ minimizes the first term inside the parenthesis on the right-hand side and leads to the relational bound \eqref{soundrel}.

\paragraph{Diffusive modes.}
Consider a diffusive mode $\omega(k) = -i D k^2 + \ldots$ with $D\in \mathbb{R}$. Consider $\im\, k = 0$ so that 
\be
- D k^2 +\ldots \leq 0
\ee
then by dividing by $k^2$ and taking the limit $k\to 0$ we obtain $D\geq 0$. The upper bound is obtained from \eqref{bestbound} at $n=2, \theta_2 = -\pi/2$, i.e. $D \leq \frac{16}{3\pi}\frac{1}{R}$. Altogether we have the two-sided bound \eqref{diffusion}. To find examples with as large as possible $DR$ we have explored the following class of test functions, 
\be
\omega(k) = A_0 + \sum_{i=1}^N A_i\sqrt{1-B_i k^2},\label{testfunction}
\ee
with $A_0$ set so that there is no gap. The largest value we found is $DR\simeq 1.22$, though we anticipate this can be improved.\footnote{With numerical parameters $N=4$, $A_1 = 0.694 + 1.929i$, $A_2 = -1.396 - 1.198i$, $A_3 = 0.429 - 1.201i$, $A_4 = -0.012 - 0.299i$, $B_1 = 0.668+0.057i$, $B_2 = 0.378-0.554i$, $B_3 = 0.109 + 0.657i$, $B_4 = 0.666 + 0.023i$, quoted to three decimal places.}
This satisfies \eqref{cond} asymptotically for large and small $k$, at each of its branch points, and has been checked numerically for $k\in \mathbb{C}$.

\section{Conclusions}

In this work we have stated a necessary condition, \eqref{cond}, for a dispersion relations $\omega(k)$ to be causal in relativistic quantum field theories, in phases where the retarded Green's function is a tempered distribution. We emphasise that \eqref{cond} is universal, independent of model details such as coupling strength.

We have proved a number of simple and universal constraints on $\omega(k)$ following from \eqref{cond}. This includes an infinite number of bounds between transport coefficients of hydrodynamics and an intrinsic microscopic scale $R$, the radius of convergence of the hydrodynamic series. Among these bounds is an upper bound on the diffusion constant, $D$ \eqref{diffusion}. There is some unexplored space $DR \in \big(1.22,\frac{16}{3\pi}\big]$; ultimately this interval can be shrunk to zero, either because the upper bound can be refined, or because there exist causal functions which have larger values of $DR$, or both.  This remains to be seen.

We note that the two-sided bounds we have obtained on hydrodynamic transport coefficients are reminiscent of causality bounds on Wilson coefficients in EFTs obtained in recent work \cite{Caron-Huot:2022ugt, CarrilloGonzalez:2022fwg}.

We have proved that causal dispersion relations  cannot be entire functions (except for the perfect fluid), and ruled out poles and essential singularities. Naively, this leaves branch point singularities in $\omega(k)$, consistent with a growing number of empirical case-by-case observations. Since in the shear channel there is only one hydrodynamic mode, the branch point implies the existence of a nonhydrodynamic shear mode. In turn, this entails a nonhydrodynamic mode in the sound channel too, by continuity at $k=0$ where the shear and sound channel degenerate.\footnote{This observation does not preclude the radius of convergence of sound channel hydrodynamic modes being set by a collision between themselves; such collisions have been observed in \cite{Novak:2018pnv}.} Thus any attempt to formulate a causal initial value problem for relativistic hydrodynamics must include nonhydrodynamic modes; the two things are inseparable. This explains the appearance of nonhydrodynamic QNMs for asymptotically AdS black branes, and nonhydrodynamic modes in phenomenological models such as RTA kinetic theory~\cite{ANDERSON1974466,Romatschke:2015gic}, MIS \cite{Israel:1976tn, Israel:1979wp}, BRSSS \cite{Baier:2007ix}, aHydro \cite{Florkowski:2010cf, Martinez:2010sc}, HJSW \cite{Heller:2014wfa}, BDNK \cite{Bemfica:2017wps, Kovtun:2019hdm, Bemfica:2019knx, Bemfica:2020zjp} and the model put forward in \cite{Heller:2021yjh}. 

\paragraph{Acknowledgements.}
\begin{acknowledgments}
We thank Christiana Pantelidou for providing the data from figure 1 of \cite{Jansen:2020hfd}. We thank Richard Davison, Luca V. Delacr\'etaz and Lorenzo Gavassino for useful correspondence.
A.\,S. acknowledges financial support from Grant No. CEX2019-000918-M funded by Ministerio de Ciencia e Innovación (MCIN)/Agencia Estatal de Investigación (AEI)/10.13039/501100011033. M.\,S. is supported by the National Science Centre, Poland, under Grants No. 2018/29/B/ST2/02457 and No. 2021/41/B/ST2/02909. B.\,W. is supported by a Royal Society University Research Fellowship and in part by the Science and Technology Facilities Council (Consolidated Grant `Exploring the Limits of the Standard Model and Beyond'). 
We would further like to thank the Isaac Newton Institute for Mathematical Sciences, Cambridge, for support and hospitality during the programme `Applicable Resurgent Asymptotics: Towards a Universal Theory'.
\end{acknowledgments}

\bibliography{causality} 

\end{document}